\documentclass[aps,prc,showpacs,superscriptaddress,twocolumn, letter]{revtex4}

\usepackage{graphicx}

\begin{document}

\title{Eccentricity fluctuations in an integrated hybrid approach: Influence
on elliptic flow}

\author{Hannah Petersen}
\affiliation{Department of Physics, Duke University, Durham, North Carolina
27708-0305, United States}

\author{Marcus Bleicher}
\affiliation{Frankfurt Institute for Advanced Studies, Johann Wolfgang
Goethe-Universit\"at, Ruth-Moufang-Str.~1,
 D-60438 Frankfurt am Main, Germany}


\begin{abstract}
The effects of initial state fluctuations on elliptic flow are investigated
within a (3+1)d Boltzmann + hydrodynamics transport approach. The spatial
eccentricity ($\epsilon_{\rm RP}$ and $\epsilon_{\rm part}$) is calculated for
initial conditions generated by a hadronic transport approach (UrQMD). Elliptic
flow results as a function of impact parameter, beam energy and transverse
momentum for two different equations of state and for averaged initial
conditions or a full event-by-event setup are presented. These investigations
allow the conclusion that in mid-central ($b=5-9$ fm) heavy ion collisions the
final elliptic flow is independent of the initial state fluctuations and the
equation of state. Furthermore, it is demonstrated that most of the $v_2$ is
build up during the hydrodynamic stage of the evolution. Therefore, the use of averaged initial profiles does not contribute to the uncertainties of the 
extraction of transport properties of hot and dense QCD matter based on viscous
hydrodynamic calculations.

\end{abstract}



\maketitle

\section{Introduction}

After the observations
at the Relativistic Heavy Ion Collider (RHIC) that the elliptic flow measurements
are consistent with ideal fluid dynamics predictions
\cite{Kolb:2000fha,Huovinen:2001cy,Csernai:2003xd} the development of numerical solutions of
viscous hydrodynamics equations is actively pursued
\cite{Song:2007fn,Song:2007ux,Luzum:2008cw,Luzum:2009sb,Song:2009rh}. The idea
of those calculations is to quantify the deviations from local thermal
equilibrium by comparisons to the available data for spectra and elliptic flow. 

Solving the differential equations of any kind of fluid dynamics prescription
implies the knowledge of the boundary conditions, i.e. the initial conditions
and the freeze-out criterion. Therefore, it is essential to investigate in a
systematic way how different initial profiles and freeze-out implementations
influence the final observable results. This article focuses on this question
concerning the initial conditions. 

The realistic situation of the collision of two nuclei suffers from many
different sources of initial state fluctuations \cite{Hirano:2009ah}. The
density profiles are not smooth, but there are peaks in the transverse and the
longitudinal direction. There are impact parameter fluctuations within one
specific centrality class leading to multiplicity fluctuations and differences in
the initial geometry \cite{Broniowski:2009fm}. Furthermore, the nuclei do not
necessarily collide in the event-plane given by the laboratory system, but
might also have a rotated reaction plane. All these effects are averaged out, if
assuming a smooth symmetric initial density profile as it is widely done by
parametrising the initial conditions for hydrodynamic calculations within a
Glauber or Color Glass Condensate (CGC) picture. To avoid potential misunderstandings, Monte-Carlo Glauber 
approaches (and CGC approaches) that produce fluctuating initial conditions are available and widely used, e.g. 
by experimental collaborations. However, usually these fluctuating initial conditions are
not used as event-by-event initial conditions for the majority of hydrodynamic simulations.
For alternative hydrodynamic approaches with fluctuating initial conditions 
see e.g. \cite{Andrade:2006yh,Werner:2009fa}.

In this article an integrated Boltzmann+hydrodynamics transport approach is applied
to the simulation of heavy ion reactions in the energy regime from $E_{\rm
lab}=2-160A$ GeV. The initial conditions are generated by the Ultra-relativistic
Quantum Molecular Dynamics (UrQMD) approach \cite{Bass:1998ca,Bleicher:1999xi}
and the above mentioned event-by-event fluctuations are further propagated in
the ideal hydrodynamic evolution employed for the hot and dense stage of the
collision. On the other hand, calculations with averaged initial conditions are
performed using the very same general setup. Comparing these calculations to the
fluctuating setup the effect on the final observable elliptic flow is
estimated.

\section{Initial Eccentricity}

Let us start by looking at the initial state eccentricity as it is given by the
UrQMD approach at the starting time $t_{\rm start} = {2R}/{\sqrt{\gamma^2 -1}}$
of the hydrodynamic evolution. For the specific values of these times for
Au+Au/Pb+Pb collisions at different beam energies see Table \ref{tab_tstart}.
This is the earliest possible transition time at which local equilibrium might
have been established after the two nuclei have passed through each other.

\begin{table}[h]
\begin{center}
\renewcommand{\arraystretch}{1.3}
\begin{tabular}{|c|c|}
\hline 
$E_{\rm lab}$ [GeV/nucleon] &  $t_{\rm start}$ [fm]\\  \hline \hline
2 & 12.417\\ 
6 & 7.169 \\
11 & 5.295 \\ 
40 & 2.830\\
160 & 1.415\\
\hline
\end{tabular}
\caption{\label{tab_tstart} Starting times of the
hydrodynamic evolution for Au+Au/Pb+Pb collisions at different beam energies. The eccentricities 
displayed in Fig. \ref{ini_ecc_exc} are also calculated at these times.}
\end{center}
\end{table}

\begin{figure}[b]
\vspace{-1cm}
\resizebox{0.5\textwidth}{!}{
\centering  
\includegraphics{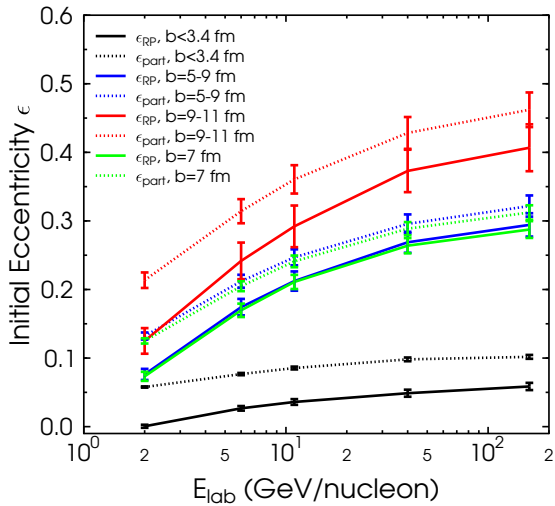}
}
\caption{(Color online) Initial eccentricity $\epsilon_{\rm RP}$ (full lines)
and $\epsilon_{\rm part}$ (dotted lines) at time $t_{\rm start}$ for Au+Au/Pb+Pb
collisions as a function of the beam energy. The results are shown for four
different centrality selections, $b < 3.4$ fm, $b=5-9$ fm, $b=9-11$ fm and fixed
impact parameter $b=7$ fm. The standard deviation $\delta \epsilon$ is depicted
as error bars to the mean value.}
\label{ini_ecc_exc}      
\end{figure}

The eccentricity quantifies the spatial anisotropy of the initial state which is
transformed via pressure gradients into a momentum space anisotropy in the
transverse plane that can be quantified by the elliptic flow coefficient. The
standard definition for the eccentricity is the reaction plane eccentricity 

\begin{equation}
\epsilon_{\rm RP} = \frac{\sigma_y^2-\sigma_x^2}{\sigma_y^2+\sigma_x^2} \quad,
\end{equation}

with $\sigma_x^2=\langle x^2 \rangle -\langle x \rangle^2$ and
$\sigma_y^2=\langle y^2 \rangle - \langle y \rangle^2$. Especially for smaller
colliding systems the so called participant eccentricity is popular 

\begin{equation}
\epsilon_{\rm part}=\frac{\sqrt{(\sigma_y^2-\sigma_x^2)^2+4
\sigma_{xy}^2}}{\sigma_y^2+\sigma_x^2} \quad ,
\end{equation}

where an additional correlation term $\sigma_{xy}=\langle xy \rangle - \langle x \rangle \langle y
\rangle $ is introduced. The participant eccentricity incorporates the fact that
the participants themselves might be rotated with respect to the reaction
plane. 
The averages $\langle \cdot \rangle$ indicate averages over all particles in one
event and the mean eccentricity is obtained by averages over many UrQMD events.
The standard deviation (shown as error bars in Fig.\ref{ini_ecc_exc}) with respect to the 
event average is $\delta \epsilon=\langle \epsilon^2 \rangle -\langle \epsilon \rangle^2$. 

\begin{figure}[t]
\resizebox{0.5\textwidth}{!}{
\centering  
\includegraphics{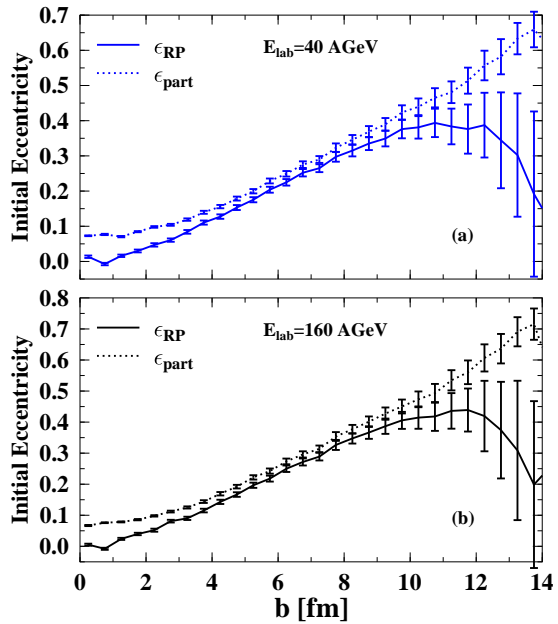}
}
\caption{(Color online) Initial eccentricity $\epsilon_{\rm RP}$ (full lines)
and $\epsilon_{\rm part}$ (dotted lines) at time $t_{\rm start}$ for Pb+Pb
collisions as a function of the impact parameter for $E_{\rm lab}=40A$ GeV (a)
and $E_{\rm lab}=160A$ GeV (b). The standard deviation $\delta \epsilon$ is
depicted as error bars to the mean value.}
\label{ini_ecc_imp}      
\end{figure}

Figs. \ref{ini_ecc_exc} and \ref{ini_ecc_imp} show the initial eccentricity in
dependence of the beam energy for different centralities and the impact
parameter dependence for two different beam energies, respectively. The full
lines depict the reaction plane eccentricity while the dotted lines refer to the
participant eccentricity. In these calculations all particles which have
suffered at least one interaction are included. This is consistent with the
particles that are taken into account for the hydrodynamic evolution where the
spectators are propagated separately in the cascade.

The initial eccentricity - mean value and standard deviation (as error bars) -
as it is shown in Fig. \ref{ini_ecc_exc} increases with the beam energy for both
definitions. For simplicity the starting times are held constant with respect to
the centrality variation. As expected the eccentricity is larger for more
peripheral collisions. The participant eccentricity is always larger than the
reaction plane eccentricity since it is calculated in the frame where the
eccentricity is the largest possible one. Also the fluctuations of the
eccentricity grow with increasing beam energy and with decreasing centrality of
the collision. The fixed impact parameter calculation for $b=7$ fm leads to very
similar results as the $b=5-9$ fm calculation for mid-central events. This hints
to the fact, that centrality fluctuations have only a minor impact on the
overall event-by-event fluctuations of the initial state.

Apart from the fact that the initial eccentricity at the highest SPS energy (see
Fig. \ref{ini_ecc_imp}) is larger than at $E_{\rm lab}=40A$ GeV the results for
both energies are pretty similar. The reaction plane eccentricity is zero for
small impact parameters while the participant eccentricity stays finite due to
the rotation to the participant plane. Going to peripheral collisions the
participant eccentricity increases almost linearly and the reaction plane
eccentricity drops down again. The fluctuations grow much more for the reaction
plane definition beyond an impact parameter of $b=10$ fm. The most interesting
observation from Fig. \ref{ini_ecc_imp} is the small difference between
$\epsilon_{\rm RP}$ and $\epsilon_{\rm part}$ and the very moderate fluctuations
for mid-central collisions in the region of $b=5-9$ fm. This behaviour is
directly reflected in the centrality dependent elliptic flow results as it is
shown in Section \ref{flow_results}. 

\section {The Hybrid Approach}

To simulate the dynamic evolution of relativistic heavy ion reactions combined
microscopic+macroscopic approaches have proven to be very successful in the
description of various observables \cite{Nonaka:2006yn,Dumitru:1999sf,Bass:2000ib,Teaney:2000cw,Teaney:2001av,Grassi:2005pm,Andrade:2005tx,Hirano:2005xf,Hirano:2007ei,Andrade:2008xh,Andrade:2008fa}. 
The approach that we are using here has recently been developed
and is based on the UrQMD hadronic transport approach including a (3+1)-dimensional 
one fluid ideal hydrodynamic evolution
\cite{Rischke:1995ir,Rischke:1995mt} for the hot and dense stage of the reaction
while the early non-equilibrium stage and the final decoupling is treated in the
hadronic cascade \cite{Petersen:2008dd,Petersen:2009vx} \footnote{The code is available as UrQMD-3.3 at http://urqmd.org}. 

For the present investigation two in principal different setups are employed,
but since they are constructed from the very same ingredients fair comparisons
can be made. The first configuration is the integrated approach where the whole
evolution from the incoming nuclei to final state particle distributions is
calculated on an event-by-event-basis. The second possibility is to stop the
UrQMD calculation at $t_{\rm start}$ and to average over many events at this
time. These averaged initial conditions are then used to calculate the
hydrodynamic evolution once and then the Cooper-Frye transition and the
subsequent hadronic rescattering is averaged over many events again to obtain
good statistics for the observables. In this second setup the spectators are not
taken into account for the further evolution. To avoid spoiling the results by
this difference in the setting we concentrate on observables at midrapidity. 

In both setups the particles in the UrQMD initial state are mapped to energy,
momentum and net baryon density distributions via three-dimensional Gaussian
distributions that represent one particle each \cite{Steinheimer:2007iy}. Two
different equations of state are used to exemplify the differences due to this
external input. One is a hadron gas equation of state (HG) with the same degrees
of freedom as in the UrQMD approach \cite{Zschiesche:2002zr}. The other one is a
bag model equation of state (BM) including a strong first order phase transition to
the quark gluon plasma with a large latent heat \cite{Rischke:1995mt}. To see if
fluctuations in the initial state affect the result differently for different
expansion dynamics during the hydrodynamic evolution these two extreme cases have been
chosen. 

The transition from the hydrodynamic evolution to the transport approach when the matter
is diluted in the late stage is treated as a gradual transition on an
approximated iso-eigentime hyper-surface (see \cite{Li:2008qm} for details).
The final rescatterings and resonance decays are taken into account in the hadronic cascade.

\section{Elliptic Flow Results}
\label{flow_results}
The elliptic flow, the second coefficient of the Fourier expansion of the
azimuthal distribution of the particles $v_2$, quantifies the momentum
anisotropy in the transverse plane
\cite{Sorge:1998mk,Ollitrault:1992bk,Bleicher:2000sx}. It is a self-quenching
effect, since the coordinate space asymmetry is transformed to a momentum space
anisotropy until the system becomes isotropic. The elliptic flow is sensitive to the
pressure gradients and therefore to the equation of state of the matter
\cite{Stoecker:1986ci,Voloshin:2008dg}. It has also been shown that the elliptic
flow is very sensitive to the shear viscosity that is present during the
evolution. 

\begin{figure}[b]
\resizebox{0.5\textwidth}{!}{
\centering  
\includegraphics{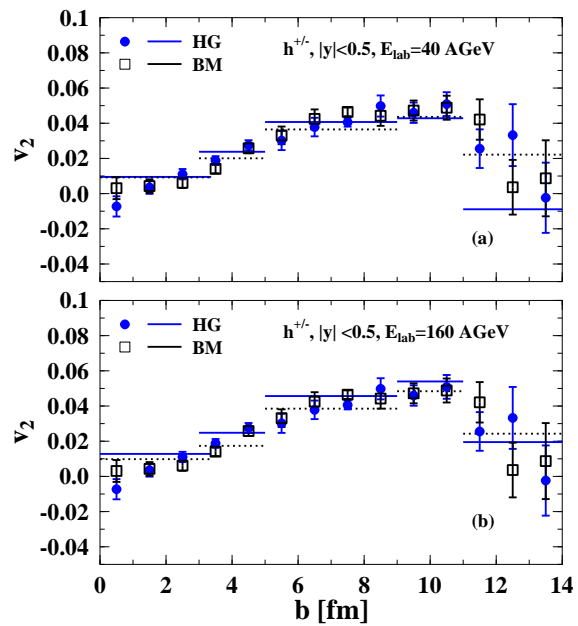}
}
\caption{(Color online) Centrality dependence of elliptic flow of charged
particles at midrapidity ($|y|<0.5$) for Pb+Pb collisions at $E_{\rm lab}=40A$
GeV (a) and $E_{\rm lab}=160A$ GeV (b). The horizontal lines indicate the
results for averaged initial conditions using the hadron gas EoS (blue full
line) and the bag model EoS (black dotted line) while the symbols (full circles for
HG-EoS and open squares for BM-EoS) depict the results for the event-by-event
calculation.}
\label{v2_imp}      
\end{figure}

In Fig. \ref{v2_imp} the centrality dependence for the charged particle elliptic
flow is shown. The upper plot (a) presents results from the hybrid model for Pb+Pb
collisions at $E_{\rm lab} =40A$ GeV while the lower plot (b) is for the highest SPS
energy. The horizontal lines depict calculations for averaged initial conditions,
while the symbols represent the full event-by-event-setup. Overall, the
differences in the integrated elliptic flow between the two different
calculations are small. Within the statistical error bars the averaged results
are in line with the results including fluctuations for the corresponding
centrality range. In accordance with the results for the impact parameter
dependence of the initial eccentricity the fluctuations get larger for
peripheral events. The highest elliptic flow values are reached for mid-central
collisions ($b=5-11$ fm). Therefore, we will refer to the impact parameter range
from $b=5-9$ fm for the following considerations. 

Furthermore, calculations for the two different equations of state are shown.
For the bag model EoS ('BM') the elliptic flow is expected to be smaller than
for the hadron gas EoS ('HG') since the speed of sound during the expansion is
smaller. This expectation is not fulfilled if one looks at the results
\cite{Petersen:2009gu}. The final elliptic flow results for different EoS are
very similar and agree within the error bars. The softer expansion seems to be
compensated by the longer lifetime of the system. 

\begin{figure}[t]
\resizebox{0.5\textwidth}{!}{
\centering  
\includegraphics{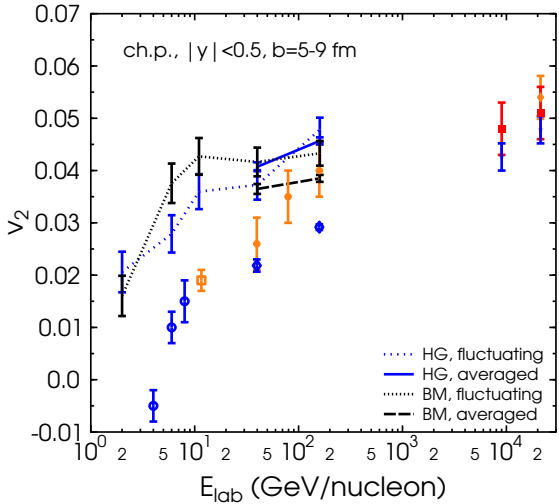}
}
\caption{(Color online) Excitation function of charged particle elliptic flow
for mid-central Au+Au/Pb+Pb collisions in comparison to the experimental data
(coloured symbols) \cite{Alt:2003ab,v2data}. Results for the hadron gas EoS with
averaged/fluctuating initial conditions are depicted as blue full/broken line
while the calculations with the bag model EoS with averaged/fluctuating initial
conditions are represented as black dashed/dotted line.}
\label{v2_exc}      
\end{figure}

The calculation of the integrated elliptic flow at midrapidity in dependence of
the beam energy (see Fig. \ref{v2_exc}) confirms the finding that the results do
not depend on the equation of state or on the initial condition setup. In the
SPS range the results of the hybrid model calculations are in a reasonable
agreement with the experimental data while at lower energies the generated
elliptic flow is too high. Here, the inclusion of a mean field in the cascade
part of the evolution helps to reproduce the data \cite{Petersen:2006vm}. 
As it has been shown in \cite{Petersen:2009vx} the elliptic flow at SPS energies 
might be affected by
the transition from hydrodynamics to the hadronic transport model and leaves
room for finite viscosities during the hydrodynamic expansion. 

\begin{figure}[t]
\resizebox{0.5\textwidth}{!}{
\centering  
\includegraphics{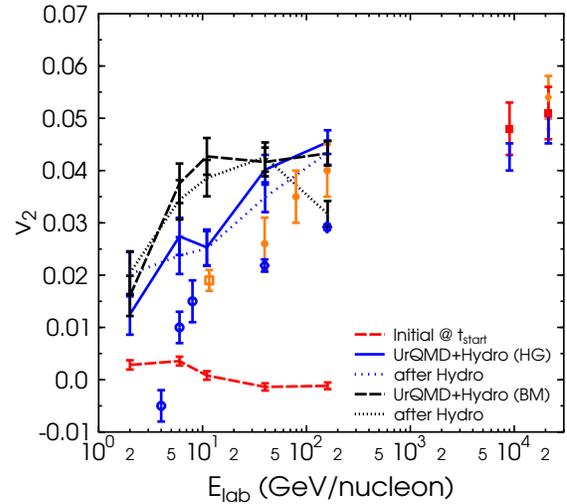}
}
\caption{(Color online) Excitation function of charged particle elliptic flow
for mid-central Au+Au/Pb+Pb collisions in comparison to the experimental data
(coloured symbols)\cite{Alt:2003ab,v2data}. The initial state elliptic flow is
shown as the red dashed line. The dotted lines refer to the end of the
hydrodynamic evolution while the full lines represents full hybrid calculations
for two different EoS.}
\label{v2_phases_exc}      
\end{figure}

One might now conclude, that the elliptic flow has to be generated in the very
early non-equilibrium stage and in the final state, if the results are not
sensitive to the EoS during the hydrodynamic expansion and do not change if the
overall setup is changed - event-by-event vs. averaged conditions. In Fig.
\ref{v2_phases_exc} the contributions to the final elliptic flow value from the
different stages of the evolution are shown. The elliptic flow in the initial
state for the hydrodynamic expansion at $t_{\rm start}$ is compatible with zero at
all beam energies. At early times, there is not enough transverse expansion to
build up elliptic flow. The dotted lines show the elliptic flow directly after
the Cooper-Frye transition at the end of the hydrodynamic expansion and at that
time the final elliptic flow is already visible. The full lines correspond to
the full calculation and indicates that only a small amount of elliptic flow is generated 
in the final and dilute stage of the collision. So, one is lead to the conclusion that the main part
of the elliptic flow is indeed generated during the ideal hydrodynamic
expansion.

\begin{figure}[t]
\resizebox{0.5\textwidth}{!}{
\centering  
\includegraphics{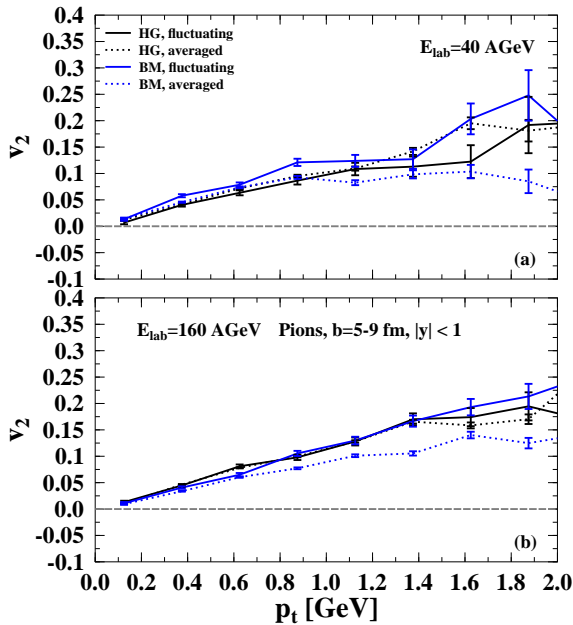}
}
\caption{(Color online) Transverse momentum dependence of elliptic flow of pions
at midrapidity in mid-central Pb+Pb collisions at SPS energies ($E_{\rm
lab}=40A$ GeV, (a) and $E_{\rm lab}=160A$ GeV, (b)). The full lines depict
results for fluctuating initial conditions while the dotted lines refer to
averaged initial conditions for two different EoS.}
\label{v2_pion_pt}      
\end{figure}

Let us now investigate the transverse momentum dependence of the elliptic flow
results. The differential flow results might be more sensitive to the different
setups. Fig. \ref{v2_pion_pt} shows the elliptic flow of pions at SPS energies
for the two different equations of state. For the hadron gas equation of state
the different setups lead to similar results again, while for the bag model the
averaged initial conditions lead to reduced elliptic flow, especially at higher
$p_t$. The Bag model equation of state seems to be more
responding to the initial energy density fluctuations. If there are fluctuations,
the transverse slices that freeze-out early and produce high transverse momentum particles have a
longer time to expand because of the higher energy density spots whereas for the
averaged conditions the whole system behaves more smoothly.

\begin{figure}[t]
\resizebox{0.5\textwidth}{!}{
\centering  
\includegraphics{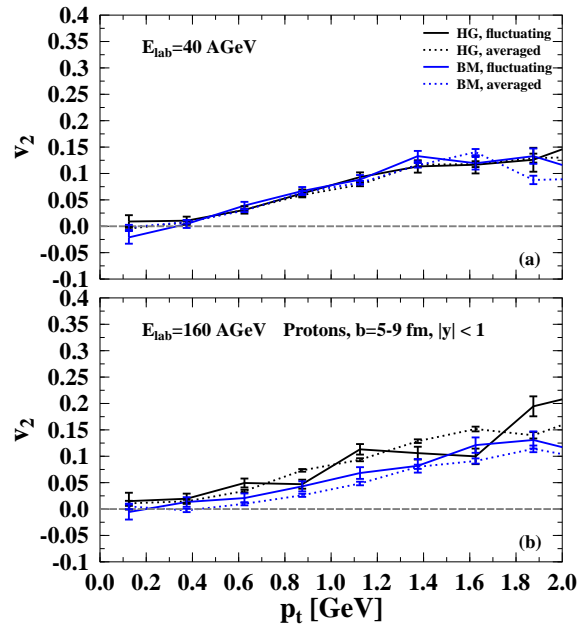}
}
\caption{(Color online) Transverse momentum dependence of elliptic flow of
protons at midrapidity in mid-central Pb+Pb collisions at SPS energies ($E_{\rm
lab}=40A$ GeV, (a) and $E_{\rm lab}=160A$ GeV, (b)). The full lines depict
results for fluctuating initial conditions while the dotted lines refer to
averaged initial conditions for two different EoS.}
\label{v2_proton_pt}      
\end{figure}

In Fig. \ref{v2_proton_pt} the transverse momentum dependence of the elliptic
flow of protons at SPS energies is shown. The protons are interesting because
they reflect the dynamics of the incoming nucleons and the finite net baryon
density. For the protons the elliptic flow results are even less sensitive to
either the EoS or the initial condition averaging than for the pions. At both
energies all calculated $v_2$ curves are compatible with each other within error bars. The
comparison to experimental data is postponed until more reliable results become
available (for pions the comparison has been published in \cite{Petersen:2009gu}).

\section{Summary and Conclusions}

The effect of initial state fluctuations on the finally measurable elliptic flow
has been studied within a (3+1)d Boltzmann + hydrodynamics transport approach. The
elliptic flow has been calculated as a function of the impact parameter, the
beam energy and the transverse momentum for two different equations of state and
for averaged initial conditions or a full event-by-event setup. These
investigations allow for the conclusion that in mid-central ($b=5-9$ fm) heavy
ion collisions the final elliptic flow reaches its maximum and the fluctuations
due to the initial state fluctuations are at a minimum. It has been confirmed
that most of the $v_2$ is build up during the hydrodynamic stage of the
evolution. The final integrated and differential elliptic flow for charged
particles at SPS energies seems to be mostly sensitive to viscosity and not so
much on the equation of state. Therefore, the use of averaged initial profiles does not contribute to the uncertainties for the extraction of transport properties
of hot and dense QCD matter based on viscous hydrodynamic calculations, while other ambiguities e.g. the contribution of the bulk viscosity \cite{Denicol:2009am} and different treatments of relaxation times for  a multi-component system \cite{Monnai:2010qp} remain under debate. Still the question if this
statement is also valid at higher RHIC energies arises, since our approach in
its present form can only be applied up to top SPS energies.

\section{Acknowledgements}
\label{ack}
We are grateful to the Center for the Scientific Computing (CSC) at Frankfurt
for the computing resources. The authors thank Dirk Rischke for providing the 
1-fluid hydrodynamics code. This work was supported by GSI and BMBF. This work was
supported by the Hessian LOEWE initiative through the Helmholtz International
Center for FAIR (HIC for FAIR). H.P. acknowledges a Feodor Lynen fellowship of
the Alexander von
Humboldt foundation. This work was supported in part by U.S. department of
Energy grant DE-FG02-05ER41367.


\end{document}